\documentclass[a4paper]{IEEEtran}

\usepackage[utf8]{inputenc} 
\usepackage[T1]{fontenc}
\usepackage{url}
\usepackage{bm}
\usepackage{ifthen}
\usepackage{cite}

\usepackage{amssymb}
\usepackage[cmex10]{amsmath} 
\usepackage{graphicx}
\usepackage{color}                             
\newcommand{\sign}[1]{{\sf sign}(#1)}
\newcommand{\tr}[1]{{\sf trace}(#1)}
\newcommand{\mmse}[1]{\eta_{MMSE}(#1)}
\newcommand{\E}[0]{{\mathbb E}}

\newtheorem{proposition}{Proposition}

\graphicspath{{./}}
\def\qed{\hfill $\Box$} 
\begin{document}
\title{Trainable ISTA for Sparse Signal Recovery}

\author{%
  \IEEEauthorblockN{Daisuke Ito\IEEEauthorrefmark{1},
  		Satoshi Takabe\IEEEauthorrefmark{1}\IEEEauthorrefmark{2},
  		and Tadashi Wadayama\IEEEauthorrefmark{1}}\\
  \IEEEauthorblockA{\IEEEauthorrefmark{1}%
		Nagoya Institute of Technology,
		Gokiso, Nagoya, Aichi 466-8555, Japan,\\
 		d.ito.480@stn.nitech.co.jp, \{s\_takabe, wadayama\}@nitech.ac.jp} \\
  \IEEEauthorblockA{\IEEEauthorrefmark{2}%
  		RIKEN Center for Advanced Intelligence Project,
  		Nihonbashi, Chuo-ku, Tokyo 103-0027, Japan
                }
\thanks{
Part of this research { was presented} 
at the IEEE International Conference of Communications 2018 (ICC2018) workshop.
}
}

\maketitle

\begin{abstract}

In the present paper, we propose a novel sparse signal recovery algorithm called the Trainable Iterative Soft Thresholding Algorithm (TISTA).
The proposed algorithm consists of two estimation units: a linear estimation unit  
and a minimum mean squared error (MMSE) estimator-based shrinkage unit.
The estimated error variance required in the MMSE shrinkage 
unit is precisely estimated from a tentative estimate of
the original signal.
The remarkable feature of the proposed scheme is 
that TISTA includes adjustable variables that control step size
and the error variance for the MMSE shrinkage.
The variables are adjusted by standard deep learning techniques. 
The number of trainable variables of TISTA is { nearly} equal to the number of iteration rounds 
and is much smaller than that of known learnable sparse signal recovery algorithms.
This feature leads to highly stable and fast training processes of TISTA.
Computer experiments show that TISTA is applicable to various classes of sensing matrices 
such as  Gaussian matrices, binary matrices, and matrices with large condition numbers.
Numerical results also demonstrate that, in many cases, TISTA provides significantly faster convergence 
than AMP and the Learned {ISTA} and { also outperforms OAMP in the NMSE performance.}
\end{abstract}

\section{Introduction}\label{sec:intro}

The basic problem setup for {\em compressed sensing}~\cite{CS1,CS2} is as follows.
A real vector $\bm{x} \in \mathbb{R}^N$ represents a sparse source signal.
It is assumed that we cannot directly observe $\bm{x}$, but 
we observe $\bm{y} = \bm{A} \bm{x} + \bm{w}$, where $\bm{A} \in \mathbb{R}^{M \times N} (N > M)$ is 
a sensing matrix and $\bm{w} \in \mathbb{R}^{M}$ is a Gaussian noise vector.
The goal is to estimate $\bm{x}$ from $\bm{y}$ as correctly as possible.

For a number of sparse reconstruction algorithms~\cite{Alg_survey}, the Lasso 
formulation \cite{LASSO} is fairly common for solving sparse signal recovery problems.
In the Lasso formulation, the original problem is recast
 as a convex optimization problem for minimizing $\frac{1}{2}  ||\bm{y} - \bm{A} \bm{x}||_2^2 + \lambda ||\bm{x}||_1$.
The regularization term $\lambda ||\bm{x}||_1$ promotes the sparseness of a reconstruction vector,
where $\lambda$ is the regularization constant.
A number of algorithms have been developed in order to solve Lasso problems efficiently~\cite{LARS}.
The Iterative Shrinkage Thresholding Algorithm (ISTA)\cite{ISTA, ISTA2}
is one of the best-known algorithms for solving the Lasso problem. 
ISTA is an iterative algorithm comprising two processes: 
a linear estimation process and a shrinkage process based on a soft thresholding function.
ISTA can be seen as a proximal gradient descent algorithm~\cite{Prox}
and can be directly derived from the Lasso formulation.

Approximate Message Passing (AMP)\cite{reBP,AMP}, which is a variant 
of approximate belief propagation, generally exhibits 
much faster convergence than the ISTA. The remarkable feature of 
AMP is that its asymptotic behavior is completely described  by
the state evolution equations~\cite{SE1}. 
AMP is derived based on the assumption that the sensing matrices consist of i.i.d. Gaussian distributed components.
Recently, Ma and Ping proposed Orthogonal AMP (OAMP) \cite{OAMP},
which can handle various classes of sensing matrices, including unitary invariant matrices. 
Rangan et al.  proposed VAMP~\cite{VAMP} for right-rotationally invariant matrices and provided a theoretical justification for its state evolution.
Independently, Takeuchi~\cite{Takeuchi} also gave a rigorous analysis for a sparse recovery algorithm for unitary invariant measurements
based on the expectation propagation framework.

The recent advent of powerful neural networks (NNs) triggered the remarkable spread of
research activities and applications on deep neural networks (DNNs)~\cite{DNN1}. 
DNN have found a number of practical applications 
such as image recognition~\cite{Image1,Image2}, speech recognition~\cite{Speech1}, 
and robotics because of their outstanding
performance compared with traditional methods. 
The advancement of DNNs has also had an impact on the design of algorithms 
for communications and signal processing~\cite{Com1,Com2}.
By unfolding an iterative process of a sparse signal recovery algorithm, 
we can obtain a {\em signal-flow graph}. The signal-flow graph
includes trainable variables that can be tuned with a supervised learning method, 
i.e., standard deep learning techniques such as 
stochastic gradient descent algorithms based on back propagation and 
mini-batches can be used to adjust the trainable variables. 
Gregor and  LeCun presented the Learned ISTA (LISTA) \cite{LISTA}, which uses 
learnable threshold variables for a shrinkage function. 
 LISTA provides a recovery performance that is 
superior to that of the original ISTA. Borgerding et al. also presented variants of AMP and VAMP with learnable capability~\cite{LAMP}  \cite{Borgerding}.

The goal of the present study is to propose a simple sparse recovery algorithm 
based on deep learning techniques. The proposed algorithm, called the {\em Trainable ISTA (TISTA)}, 
borrows the basic structure of ISTA, and adopts the estimator of the squared error between 
true signals and tentative estimations, i.e., the {\em error variance estimator}, from OAMP { \cite{OAMP}}. 
Thus, TISTA 
consists of the three parts: 
a linear estimator, a minimum mean squared error (MMSE) 
estimator-based shrinkage function, and the above-mentioned error variance estimator.
The linear estimator of TISTA includes trainable variables that can be adjusted via deep learning techniques.
Zhang and Ghanem \cite{Zhang} proposed ISTA-Net, which is also an ISTA-based algorithm 
with learnable capability. The notable difference between ISTA-Net and TISTA is that 
TISTA uses an error variance estimator, which { significantly} improves the speed of convergence.

\section{Brief review of known recovery algorithms}

As preparation for describing the details of the proposed algorithm, 
several known sparse recovery algorithms are briefly reviewed in this section.
In the following, the observation vector is assumed to be
$\bm{y} = \bm{A} \bm{x} + \bm{w}$, where $\bm{A} \in \mathbb{R}^{M \times N} (N > M)$ 
and $\bm{x}\in \mathbb{R}^N$. Each entry of the additive noise vector $\bm{w} \in \mathbb{R}^{M}$ 
follows a zero-mean Gaussian distribution with variance $\sigma^2$.

\subsection{ISTA}

The ISTA is a well-known sparse recovery algorithm \cite{ISTA}
defined by the following simple recursion:
\begin{eqnarray}
\bm{r}_t &=& \bm{s}_t + \beta \bm{A}^T(\bm{y} - \bm{A} \bm{s}_t)\\
\bm{s}_{t + 1} &=& \eta(\bm{r}_t; \tau),
\end{eqnarray}
where $\beta \in \mathbb R$ represents the step size, and 
$\eta(\cdot; \cdot): \mathbb R^n \rightarrow \mathbb R^n$ is the soft thresholding function defined by 
\[
\eta (\bm{r}; \tau) = (\tilde \eta(r_1; \tau), \ldots,  \tilde \eta(r_n; \tau)),
\]
where $\tilde \eta (\cdot; \cdot): \mathbb R \rightarrow \mathbb R$ is given by
\begin{equation}
\tilde \eta (r; \tau)  = \sign{r} \max \{ |r| - \tau, 0 \}.	
\end{equation}
The parameter $\tau \in \mathbb R (\tau > 0)$ indicates the threshold value.
After $T$-iterations, the estimate $\hat{\bm {x}} = \bm{s}_T$ of the original sparse signal $\bm{x}$ is obtained.
The initial value is assumed to be $\bm{s}_0 = \bm{0}$. In order to have convergence, 
the {step size $\beta$} should be carefully determined \cite{ISTA}. Several accelerated methods
for ISTA using a momentum term, such as the Fast ISTA (FISTA),  
have been proposed~\cite{TwIST,  FISTA}.
Since the proximal operator of the $\ell_1$-regularization term $||\bm{x}||_1$ is the soft thresholding function, 
the ISTA can be seen as a proximal gradient descent algorithm~\cite{Alg_survey}.

\subsection{AMP}

AMP\cite{AMP} is defined by the following recursion:
\begin{eqnarray}
\bm{r}_t &=& \bm{y} - \bm{A} \bm{s}_t + b_t \bm{r}_{t-1},\label{onsager}\\ \label{shrinkage}
\bm{s}_{t + 1} &=& \eta(\bm{s}_t + \bm{A}^T \bm{r}_t; \tau_t), \\ \label{tau_est_amp}
b_t &=& \frac{1}{M} || \bm{s}_t ||_0,\quad \tau_t = \frac{\theta}{\sqrt{M}} || \bm{r}_t ||_2
\end{eqnarray}
and provides the final estimate $\hat{\bm{x}} = \bm{s}_T$. 
Each entry of the sensing matrix $\bm{A}$ is assumed to be generated according to the Gaussian distribution 
$\mathcal{N}(0, 1/M)$, i.e., a Gaussian distribution with mean zero and variance $1/M$.
At a glance, the recursive formula of AMP appears similar to that of ISTA, but there are several critical differences.
Due to the {\em Onsager correction term} $b_t \bm{r}_{t-1}$ in (\ref{onsager}),
the output of the linear estimator becomes statistically decoupled, 
and an error between each output signal from the linear estimator and the true signal 
behaves as a white Gaussian random variable in the large system limit. This enables us to use a 
scalar recursion called the {\em state evolution} to track the evolution of the error variances.

Another difference between ISTA and AMP is the estimator of $\tau_t$ in (\ref{tau_est_amp}), which 
is used as the threshold value for the shrinkage function (\ref{shrinkage}).
In \cite{AMP}, it was reported that AMP exhibits much faster 
convergence than  ISTA if the sensing matrix satisfies the above condition.
On the other hand, 
AMP cannot provide excellent recovery performance 
for sensing matrices violating the above condition 
such as non-Gaussian sensing matrices, Gaussian matrices with large variance, 
Gaussian matrices with nonzero means, and matrices with large condition numbers~\cite{IC1}. 

\subsection{OAMP}
OAMP\cite{OAMP} is defined by the following recursive formula:
\begin{eqnarray}
\bm{r}_t &=& \bm{s}_t + \bm{W}(\bm{y} - \bm{A} \bm{s}_t), \label{oamp_r}\\
\bm{s}_{t + 1} &=& \eta_{\text{df}}(\bm{r}_t; \tau_t), \label{oamp_s}\\
v_t^2 &=& \max \left\{ \frac{||\bm{y} - \bm{A} \bm{s}_t||_2^2 - M \sigma^2}{\tr{\bm{A}^T \bm{A}}}, \epsilon \right\}, \label{v_est} \\ \label{tau_est_oamp}
\tau_t^2 &=& \frac{1}{N} \tr{\bm{B} \bm{B}^T} v_t^2 + \frac{1}{N} \tr{\bm{W} \bm{W}^T} \sigma^2, 
\end{eqnarray}
for $t = 0,1,2, \ldots, T-1$. The matrix $\bm{B}$ is given by $\bm{B} = \bm{I} - \bm{W} \bm{A}$.
To be precise, the estimator equations on $v_t^2$ (\ref{v_est}) and 
$\tau_t^2$ (\ref{tau_est_oamp}) (also presented in \cite{EGAMP}) 
are not part of OAMP (for example, we can use the state evolution 
to provide $v_t^2$ and $\tau_t^2$), but these estimators are used for 
numerical evaluation in \cite{OAMP}.
The matrix $\bm{W}$ in linear estimator (\ref{oamp_r}) can be chosen from the transpose of $\bm{A}$, 
the pseudo inverse of $\bm{A}$, and the LMMSE matrix.
The nonlinear estimation unit (\ref{oamp_s}) consists of a {\em divergence-free function} 
$\eta_{\text{df}}$ that replaces the Onsager correction term.
It is proved in \cite{OAMP} that the estimation errors of linear estimator (\ref{oamp_r})  and non-linear estimator (\ref{oamp_s}) are
 statistically orthogonal if a sensing matrix is i.i.d. Gaussian or unitary invariant. This provides a justification for the state evolution of OAMP.

\section{Details of  TISTA}

This section describes the details of TISTA and its training process.

\subsection{MMSE estimator for an additive Gaussian noise channel}

Let $X$ be a real-valued random variable with probability density function (PDF) $P_X(\cdot)$.
We assume an additive Gaussian noise channel defined by 
$
  Y = X + N,
$
where $Y$ represents a real-valued random variable as well.
The random variable $N$ is a Gaussian random variable 
with mean $0$ and variance $\sigma^2$. 
Consider the situation in which a receiver can observe $Y$ and we wish to estimate the value of $X$.

The MMSE estimator $\mmse{y}$  is  defined by
\begin{equation}
\mmse{y} = \E [X | y],
\end{equation}
where $\E [X | y]$ is the conditional expectation given by
\begin{equation}
	\E [X| y] = \int_{-\infty}^{\infty} x P(x|y) dx.
\end{equation}
The posterior PDF $P(x|y)$ is given by Bayes' Theorem:
\begin{equation}
	P_{X|Y}(x|y) = \frac{P_X(x) P_{Y|X}(y|x)}{P_Y(y)},
\end{equation}
where the conditional PDF is Gaussian:
\begin{equation}
P_{Y|X}	(y|x) = \frac{1}{\sqrt{2 \pi \sigma^2}}\exp \left(\frac{-(y-x)^2}{2 \sigma^2}  \right).
\end{equation}

In the case of the Bernoulli-Gaussian prior, $P_X(x)$ is given by
\begin{equation}\label{Bprior}
	P_X(x) = (1 - p) \delta(x) +
	\frac{p}{\sqrt{2 \pi \alpha^2}} 
	\exp \left(-\frac{x^2}{2 \alpha^2} \right),
\end{equation}
where $p$ represents the probability such that a nonzero element occurs.
The function $\delta(\cdot)$ is Dirac's delta function.
In this case,  a nonzero element follows the Gaussian PDF with mean $0$ and variance 
$\alpha^2$.
The MMSE estimator for the Bernoulli-Gaussian prior can be easily derived \cite{Gribonval} using Stein's formula:
\begin{equation}
 \mmse{y; \sigma^2} = y + \sigma^2 \frac{d}{dy} \ln P_Y(y)
\end{equation}
and we have 
\begin{equation} \label{mmse_est}
  \mmse{y; \sigma^2} = \left(\frac{y \alpha^2}{\xi}\right) 
  \frac{p F(y;\xi)}{(1-p) F(y; \sigma^2)  
  + p F(y; \xi)},
\end{equation}
where $\xi = \alpha^2 + \sigma^2$ and 
\begin{equation}
	F(z; v) = \frac{1}{\sqrt{2 \pi v}} \exp \left( \frac{-z^2}{2 v} \right).
\end{equation}

\begin{figure}[bt]
        \begin{center}
          \includegraphics[width=0.95\linewidth]{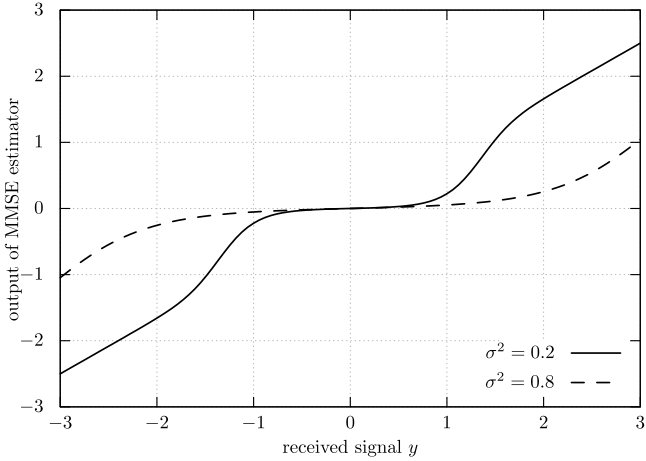}
          \caption{Plots of $\eta_{MMSE}$ as a function of a received signal $y$ ($\alpha^2 = 1$, $\sigma^2 = 0.2, 0.8$, $p=0.1$).}
          \label{mmse_graph}
        \end{center}
\end{figure}

For example, Fig. \ref{mmse_graph} shows the shapes of
$\eta_{MMSE}(y; \sigma^2)$ as a function of a received signal $y$
for $\sigma^2 = 0.2, 0.8$ .
The shapes can be observed to resemble those of the soft thresholding function but 
the function is differentiable everywhere with respect to $y$.

Let us consider another setting.
If each sparse component takes a value in a finite discrete set
$
	S = \{s_1, \ldots, s_M \} (s_i \in \mathbb{R})
$
uniformly at random, then the corresponding prior becomes
\begin{equation}
	P_X(x) = (1- p) \delta(x) + p \sum_{s \in S} \frac{1}{M} \delta (x - s),
\end{equation}
and we have the MMSE estimator 
\begin{equation}
  \mmse{y; \sigma^2}
  =	 \frac{p \sum_s s F(s; \sigma^2) }
{(1-p) M F(0; \sigma^2) + p \sum_{s} F(s; \sigma^2)}.
\end{equation}

These MMSE estimators are going to be used as a building block of the
TISTA to be presented in the next subsection.

\subsection{Recursive formula for TISTA}

We assume that the sensing matrix $\bm{A} \in \mathbb R^{M \times N}$ is a 
full-rank matrix.
The recursive formula of TISTA is summarized as follows:
\begin{eqnarray}
\bm{r}_t &=& \bm{s}_t + \gamma_t \bm{W} (\bm{y} - \bm{A} \bm{s}_t), \label{rt} \label{lmse}\\
\bm{s}_{t+1} &=& \mmse {\bm{r}_t; \tau_t^2},  \label{st} \label{mmse}\\
v_t^2 &=& \max \left\{ \frac{||\bm{y} - \bm{A} \bm{s}_t||_2^2 - M \sigma^2}{\tr{\bm{A}^T \bm{A}}}, \epsilon \right\}, \label{v_t}\\ \label{tau_est}
\tau_t^2 &=& \frac{v_t^2}{N} (N + (\gamma^2_t  - 2\gamma_t)  M   )    \nonumber\\
&+& \frac{\gamma_t^2\sigma^2}{N} 
\tr{\bm{W} \bm{W}^T}, 
\end{eqnarray}
where the matrix $\bm{W} = \bm{A}^T(\bm{A} \bm{A}^T)^{-1}$
is the pseudo inverse matrix 
of the sensing matrix $\bm{A}$.
The initial condition is $\bm{s}_0 = 0$, and the final estimate 
is given by $\hat{\bm{x}} = \bm{s}_T$.   
The scalar variables $\gamma_t \in \mathbb{R} (t = 0,1,\ldots, T-1)$ are learnable 
variables that are tuned in a training process.
The number of learnable variables is thus $T$, which is much smaller than 
those of LISTA~\cite{LISTA} and LAMP~\cite{LAMP}. 
In addition { to the step size parameters $\{\gamma_t\}_{t=0}^{T-1}$, one can also optimize} parameters $p$ and $\alpha$
 in the MMSE estimator (\ref{mmse_est}) especially for nonsynthetic signals or real data.
We assume that they are constant among iterations in TISTA for simplicity.
The number of { the trainable parameters in this case} is thus $T+2$.

An appropriate MMSE shrinkage (\ref{mmse}) is chosen according 
to the prior distribution of the original signal $\bm{x}$.
Note that the MMSE shrinkage is also used in \cite{LAMP}.
The real constant $\epsilon$ is a sufficiently small value, e.g.,  $\epsilon = 10^{-9}$.
The max operator in (\ref{v_t}) is used to prevent the estimate of the variance 
from being non-positive.
The learnable variables $\gamma_t$ in (\ref{rt}) provide appropriate step sizes and 
control for the variance of the MMSE shrinkage.

The true error variances $\bar \tau_t^2$ and $\bar v_t^2$ are defined by
\begin{equation}
\bar \tau_t^2 = \frac{\E [|| \bm{r}_t - \bm{x} ||_2^2]}{N}, \quad
\bar v_t^2 = \frac{\E [|| \bm{s}_t - \bm{x} ||_2^2]}{N}.
\end{equation}
These error variances should be estimated as correctly as possible in a sparse recovery process 
because the MMSE shrinkage unit 
(\ref{mmse}) requires knowing $\bar \tau_t^2$.
As in the case of OAMP \cite{OAMP}, we make the following assumptions 
on the residual errors in order to derive an error variance estimator.

The first assumption is that $\bm{r}_t - \bm{x}$ consists of i.i.d. zero-mean Gaussian entries.
Based on this assumption, each entry of the output from the linear estimator (\ref{rt}) can be seen as an observation obtained from a {virtual additive Gaussian noise channel} 
with the noise variance $\bar \tau^2$.
This justifies the use of the shrinkage function based on the MMSE estimator (\ref{mmse}) 
with $\bar \tau^2$. Another assumption is that 
$\bm{s}_{t} - \bm{x}$ consists of zero-mean i.i.d. entries and satisfies 
$\E[(\bm{s}_t - \bm{x})^T \bm{A}^T \bm{w}] = \E[(\bm{s}_t - \bm{x})^T \bm{W} \bm{w}] = 0$ for any $t$.

The error variance estimator for $\bar v_t^2$ (\ref{v_t}) is the same as that of OAMP \cite{OAMP},
and its justification comes from the following proposition.
\begin{proposition}
If each entry of $\bm{s}_t - \bm{x}$ is i.i.d. with mean zero and
$\E [(\bm{s}_t - \bm{x})^T  \bm{A}^T \bm{w}] = 0$ is satisfied, then 
\begin{equation}\label{vtrue}
  \bar v_t^2 =  \frac{\E[||\bm{y} - \bm{A} \bm{s}_t||_2^2] - M \sigma^2}{ \tr{\bm{A}^T \bm{A}} } 
\end{equation}
holds.
\end{proposition}
(Proof) From the right-hand side of (\ref{vtrue}), we have
\begin{align*}
  &\frac{\E[||\bm{y} - \bm{A} \bm{s}_t||_2^2] - M \sigma^2}{ \tr{\bm{A}^T \bm{A}} } \nonumber\\\nonumber
  =&   \frac{\E[||\bm{A} \bm{x} + \bm{w} - \bm{A} \bm{s}_t||_2^2] - M \sigma^2}{ \tr{\bm{A}^T \bm{A}} }  \\ \nonumber
  =&   \frac{\E[||\bm{A}(\bm{x} - \bm{s}_t) + \bm{w}||_2^2] - \E[\bm{w}^T \bm{w}]  }{ \tr{\bm{A}^T \bm{A}} }  \\ \nonumber
  =&   \frac{\E[(\bm{A}(\bm{x} \!-\! \bm{s}_t))^T\! \bm{A} (\bm{x}\!-\! \bm{s}_t)  \!+\! (\bm{A} (\bm{x}\!-\! \bm{s}_t))^T \bm{w} ]   }{ \tr{\bm{A}^T \bm{A}} }  \\ \nonumber
  =&   \frac{\E[  (\bm{x} - \bm{s}_t)^T \bm{A}^T \bm{A} (\bm{x}- \bm{s}_t)]}{ \tr{\bm{A}^T \bm{A}} } \\ \nonumber
  =&  \frac{1}{N} \tr{\bm{A}^T \bm{A}} \E[||\bm{s}_t - \bm{x}||_2^2] \frac{1}{\tr{\bm{A}^T \bm{A}}} \\ 
  =&  \frac{1}{N}  \E[||\bm{s}_t - \bm{x}||_2^2] = v_t^2.
\end{align*}
\hfill\qed

The justification of the error variance estimator (\ref{tau_est}) for $\bar \tau_t^2$ is also provided 
by the following proposition.
\begin{proposition}
If each entry of $\bm{s}_t - \bm{x}$ is i.i.d. with mean zero and
$\E [(\bm{s}_t - \bm{x})^T  \bm{W} \bm{w}] = 0$ is satisfied, then
\begin{eqnarray} \nonumber
  \bar \tau_t^2 &=& \frac {\bar v_t^2}{N}  (N  - 2\gamma_t  \tr {\bm{Z}} + \gamma^2_t \tr {\bm{Z} \bm{Z}^T} )  \\
  &+& \frac {\gamma_t^2 \sigma^2} N \tr{\bm{W} \bm{W}^T}
\end{eqnarray}
holds, where $\bm{Z} = \bm{W} \bm{A}$.
\end{proposition}
(Proof) The residual error $\bm{r}_t- \bm{x}$ can be rewritten as
\begin{eqnarray}  \nonumber
\bm{r}_t - \bm{x} &=& \bm{s}_t + \gamma_t \bm{W} (\bm{y} - \bm{A} \bm{s}_t) - \bm{x}\\  \nonumber
&=& \bm{s}_t + \gamma_t \bm{W} (\bm{A} \bm{x} + \bm{w})- \gamma_t \bm{W} \bm{A} \bm{s}_t - \bm{x}  \\ \nonumber
&=& (\bm{I} -\gamma_t \bm{Z})(\bm{s}_t - \bm{x}) + \gamma_t \bm{W} \bm{w}.
\end{eqnarray}
From the definition $\bar \tau_t^2$,  we have 
\begin{align*} \nonumber
\bar \tau_t^2 &=  \frac 1 N \E [||(\bm{I} - \gamma_t \bm{Z}) (\bm{s}_t - \bm{x})+ \gamma_t \bm{W} \bm{w}||_2^2 ]  	 \\ \nonumber
&=  \frac 1 N \E [ (\bm{s}_t - \bm{x})^T (\bm{I} - \gamma_t \bm{Z}) (\bm{I} - \gamma_t \bm{Z})^T (\bm{s}_t - \bm{x})]  \\ \nonumber
&+   \frac {\gamma_t^2}{N} \E [\bm{w}^T \bm{W}^T \bm{W} \bm{w}]   
+  \frac {2 \gamma_t } {N} \E [(\bm{s}_t - \bm{x})^T (I \!-\! \gamma_t \bm{Z})^T \bm{W} \bm{w}] \\ \nonumber
&=  \frac 1 N \tr{(\bm{I} - \gamma_t \bm{Z}) (\bm{I} - \gamma_t \bm{Z})^T} \bar v_t^2  \\ \nonumber
&+   \frac {\gamma_t^2}{N} \tr{\bm{W} \bm{W}^T}\sigma^2  
+  \frac {2  (\gamma_t - \gamma_t^2)} {N} \E [(\bm{s}_t - \bm{x})^T  \bm{W} \bm{w}].
\end{align*}
The last term vanishes due to the assumption 
$
\E [(\bm{s}_t - \bm{x})^T  \bm{W} \bm{w}] = 0,
$
and the first term can be rewritten as 
\begin{eqnarray} \nonumber
&&\hspace{-1.5cm} \tr{ (\bm{I} - \gamma_t \bm{Z}) (\bm{I} - \gamma_t \bm{Z})^T	 } \\ \nonumber
	&=& 	\sum_{i,j: i \ne j} (\gamma_t \bm{Z}_{i,j})^2 + \sum_{i}  (1 - \gamma_t \bm{Z}_{i,i})^2 \\ \nonumber
	&=& \gamma^2_t \sum_{i,j: i \ne j} \bm{Z}_{i,j}^2 + \sum_{i}  (1 - 2\gamma_t \bm{Z}_{i,i} + \gamma^2_t \bm{Z}_{i,i}^2 ) \\
	&=& N  - 2\gamma_t  \tr {\bm{Z}} + \gamma^2_t \tr {\bm{Z} \bm{Z}^T}.
\end{eqnarray}
The proposition is thus proved. \hfill \qed\\
The identity $\tr {\bm{Z}}  =\tr {\bm{Z} \bm{Z}^T} = M$ holds because $\bm{A}$ and $\bm{Z}$ have full rank.
Combining this identity, 
we have the estimation formula (\ref{tau_est}) for $\tau_t^2$.

These error variance estimators (\ref{v_t}) and (\ref{tau_est}) 
play a crucial role in providing appropriate variance estimates required for the MMSE shrinkage.
Since the validity of these assumptions on the residual errors cannot be proved,
it will be experimentally confirmed in the next section.
Moreover, note that the TISTA recursive formula {does not include either}
an Onsager correction term or a divergence-free function. 
Thus, we cannot expect stochastic orthogonality guaranteed in OAMP in a process of TISTA. 
This means that the state evolution cannot be used to analyze the asymptotic performance of TISTA. 

\subsection{Time complexity and number of trainable variables}

For treating a large-scale problem, a sparse recovery algorithm should require low 
computational complexity for each iteration.
The time complexity required for evaluating the recursive formula 
of TISTA per iteration is $O(N^2)$, which is the same time complexity as those of ISTA and AMP, which means that the TISTA has sufficient scalability for large problems.
The evaluation of the matrix-vector products $\bm{A} \bm{s}_t$ and $\bm{W} (\bm{y} - \bm{A} \bm{s}_t)$ 
requires $O(N^2)$ time, which is dominant in an iteration.
The evaluation of the scalar constants $\tr{\bm{A}^T \bm{A}}$ and  $\tr{\bm{W} \bm{W}^T}$
requires $O(N^2)$ time.
Although computation of the pseudo inverse of $\bm{A}$ requires
$O(N^3)$ time, it can be pre-computed only once in advance.

\begin{table}
\begin{center}
\caption{Numbers of trainable variables in the $T$-round process}
\label{num_vars}
  \begin{tabular}{c||c|c|c} \hline \hline
     & TISTA & LISTA & LAMP\\ \hline
    \# of params & $T+2$ & $T(N^2 + MN + 1)$ & $T(NM + 2)$ \\ \hline
  \end{tabular}
\end{center}
\end{table}

{ Since the $t'$-th round of TISTA contains 
only trainable variables $\{\gamma_t\}_{t=0}^{t'-1}$  (or $\{\gamma_t\}_{t=0}^{t'-1}$, $\alpha$ and $p$),}
the total number of trainable variables is $T$ (or $T+2$) { for TISTA with $T$ iteration rounds.}
On the other hand,  LISTA and LAMP require
$N^2 + MN + 1$ and  $NM + 2$ trainable variables for 
each round, respectively.
Table \ref{num_vars} summarizes 
the required numbers of trainable variables in $T$ rounds.
TISTA requires the least trainable variables among them, and
the number of trainable variables of  TISTA is independent of the 
system size, i.e., $N$ and $M$.
This is an advantageous feature for large-scale problems.
The number of trainable variables 
also affects the stability and speed of convergence in training processes.

\subsection{Incremental training for TISTA}

\begin{figure}[bt]
        \begin{center}
          \includegraphics[scale=0.3]{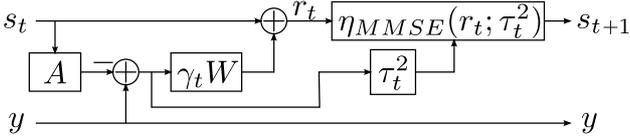}
          \caption{{Schematic diagram of the} $t$-th iteration of TISTA with learnable variable $\gamma_t$.}
          \label{fig:tista}
        \end{center}
\end{figure}

In order to achieve reasonable recovery performance, 
the trainable variables { $\{\gamma_t \}_{t=0}^{T-1}$} should be appropriately adjusted.
By unfolding the recursive formula of  TISTA, we immediately have a signal-flow graph which is similar to a multi-layer feedforward neural network.
Figure \ref{fig:tista} depicts a unit of the signal-flow graph corresponding to the 
$t$-th iteration of  TISTA, and we can stack the units to compose a whole signal-flow graph.
Here, we follow a standard recipe of deep learning techniques; namely, we apply mini-batch training with a stochastic gradient descent algorithm to the signal-flow
graph of TISTA. 
Based on several experiments, we found that the following {\em incremental training} 
is considerably effective for learning appropriate values that provide superior performance. 
This is because the {\em vanishing gradient problem} makes one-shot training for the whole network difficult.
The incremental training discussed below can reduce the effect of the vanishing gradient. 

The training data consists of a number of randomly generated pairs $(\bm{x}, \bm{y})$, where $\bm{y} = \bm{A} \bm{x} 
+ \bm{w}$. The sample $\bm{x}$ follows the prior distribution $P_X(\bm{x})$ 
and the observation noise $\bm{w}$ is an i.i.d. Gaussian random vector.
The entire set of training data is divided into mini-batches to be used in a stochastic gradient descent algorithm
such as SGD, RMSprop, or Adam.

In the $t$-th round of the incremental training (referred to as a {\em generation}),  an optimizer attempts to minimize
$
	\E [||\bm{s}_{t} - \bm{x}||_2^2]
$
{ by tuning $\{\gamma_{t'}\}_{t'=0}^{t-1}$} (and possibly $\alpha$ and $p$). 
The number of mini-batches used in the $t$-th generation 
is denoted by $D$. After processing $D$ mini-batches, 
the objective function of the optimizer 
is changed to $\E [||\bm{s}_{t+1} - \bm{x}||_2^2]$. Namely, after training the first to $t$-th layers, 
a new $t+1$ layer is appended to the network, and the entire network is trained again for $D$ mini-batches.
Although the objective function is changed, 
the values of the variables $\gamma_0,\ldots, \gamma_{t-1}$ of the previous generation are taken as the initial values in the optimization process for the new generation. 
In summary,  the incremental training updates the variables $\gamma_t$ in a sequential manner from 
the first layer to the last layer.

\section{Performance evaluation}\label{sec_eval}
{ In this section, the sparse recovery performance of TISTA is evaluated by computer experiments.}

\subsection{Details of experiments}

The basic conditions for the computer experiments { shown in this section} are summarized as follows.
Each component of the sparse signal $\bm{x}$ is assumed to be a realization of an 
i.i.d. random variable following the Bernoulli-Gaussian PDF (\ref{Bprior})
with $p = 0.1, \alpha^2 = 1$. 
The Bernoulli-Gaussian PDF is often assumed as a benchmark setting in related researches\cite{LAMP, Borgerding}.
We thus
use the MMSE estimator (\ref{mmse}) for the Bernoulli-Gaussian prior.
Each component of the noise vector $\bm{w}$ follows the zero-mean Gaussian PDF
with variance $\sigma^2$.
The signal-to-noise ratio (SNR) of the system is
defined as 
\begin{equation}
SNR = \frac{\E [||\bm{A} \bm{x}||_2^2]}{\E [||\bm{w}||_2^2]}.
\end{equation}
The size of the mini-batch is set to $1000$, and $D = 200$ mini-batches are 
allocated for each generation. We used the Adam optimizer \cite{Adam}.
{The learning rate of the optimizer is set to $4.0\times 10^{-2}$ in the first 10 iterations and
$8.0\times 10^{-4}$ in the remaining iterations.}
The experimental system was implemented in TensorFlow~\cite{tens} and PyTorch~\cite{Pytorch}.
For comparison purposes, we will include the NMSE performances 
of AMP and other algorithms in the following subsections.
The {hyperparameter} $\theta$ used in AMP is set to 
$\theta = 1.14$.
We used an implementation of LISTA \cite{git} by the authors of \cite{LAMP}.

\subsection{IID Gaussian matrix with small variance}

Here, we consider { the conventional setting  for compressed sensing in which AMP successfully indicates convergence.
 The trainable parameters of TISTA in this subsection are $\{\gamma_t\}_{t=0}^{T-1}$, $\alpha$, and $p$.}

\subsubsection{Comparison with AMP and other algorithms}

This subsection describes the case in which $\bm{A}_{i,j} \sim \mathcal{N}(0, 1/M)$, i.e.,
each component of the sensing matrix $\bm{A}$ obeys a zero-mean Gaussian distribution 
with variance $1/M$. Note that AMP is designed for this matrix ensemble.
The dimensions of the sensing matrices are set to be $N = 500, M = 250$.

\begin{figure}[bt]
        \begin{center}
          \includegraphics[width=0.95\linewidth]{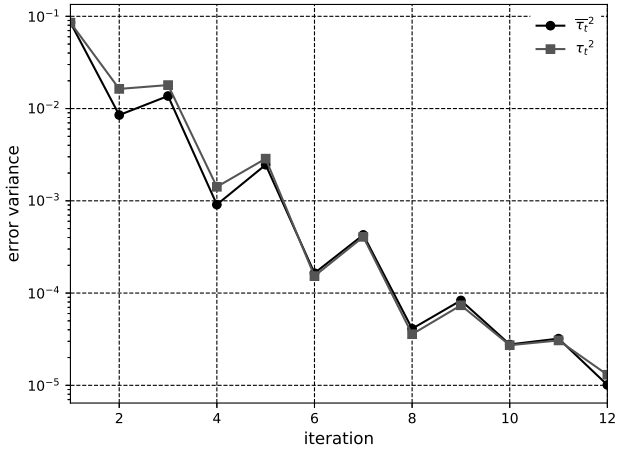}
          \caption{Estimate $\bar{\tau}^2$ and 
          the true error variance $\tau^2$;
          {$\bm{A}_{i,j} \sim \mathcal{N}(0, 1/M), N = 500, M = 250$, SNR $= 40$ dB}.
}
          \label{fig:tau}
        \end{center}
\end{figure}

\begin{figure}
        \begin{center}
          \includegraphics[width=0.95\linewidth]{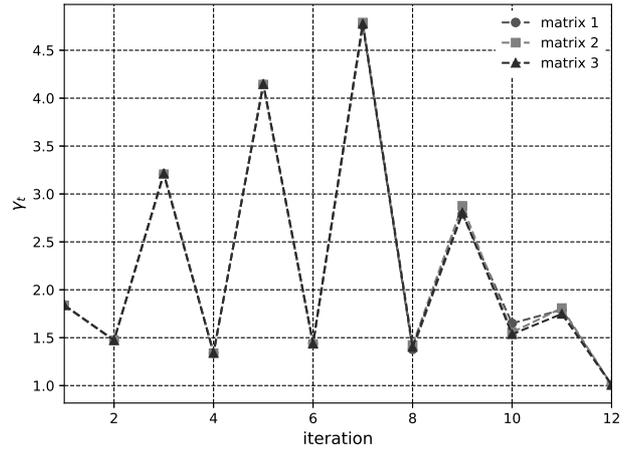}
          \caption{Three sequences of learned variables $\gamma_t$; $\bm{A}_{i,j} \sim \mathcal{N}(0, 1/M), N = 500, M = 250, p = 0.1$, SNR $= 40$ dB. }
          \label{A_variance:1_M:g}
        \end{center}
\end{figure}

Figure \ref{fig:tau} shows the estimate $\tau^2$ by (\ref{tau_est}) and
the empirically estimated values of the true error variance $\bar{\tau}^2$. The estimator ${\tau}^2$ provides accurate estimations and justifies 
the use of (\ref{v_t}) and  (\ref{tau_est}) and our assumptions on the residual errors. 
{We find that the error variance does not monotonically decrease.
Because the residual error depends on the trainable parameters $\{\gamma_t\}_{t=0}^{T-1}$,
the zigzag shape of $\gamma_t$'s (see Fig. \ref{A_variance:1_M:g}) may affect the shapes of ${\tau}^2$ and $\bar{\tau}^2$.
In spite of this nontrivial tendency, the residual error decreases rapidly indicating a successful signal recovery.}

\begin{figure}[bt]
        \begin{center}
          \includegraphics[width=0.95\linewidth]{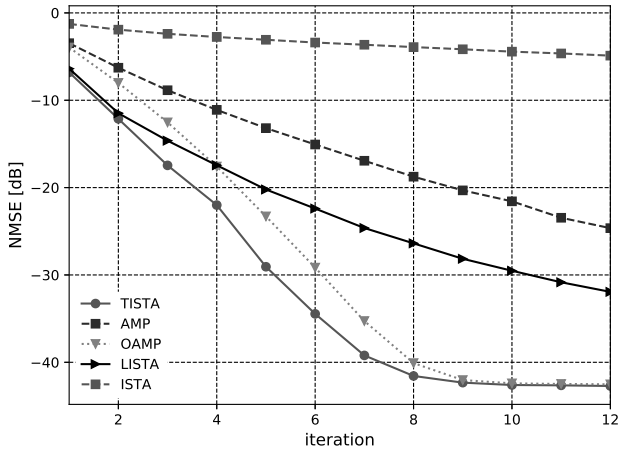}
          \caption{NMSE of TISTA and other algorithms;  $\bm{A}_{i,j} \sim \mathcal{N}(0, 1/M), N = 500, M = 250$, SNR $= 40$ dB.
           Condition $\bm{A}_{i,j} \sim \mathcal{N}(0, 1/M)$ is required for AMP to converge.}
          \label{A_variance:1_M}
        \end{center}
\end{figure}

Figure \ref{A_variance:1_M} presents the average normalized MSE (NMSE) 
of TISTA, ISTA, LISTA, AMP, and OAMP as functions of iteration when SNR $= 40$ dB.
The NMSE  is defined by 
\begin{equation}
NMSE = 10 \log_{10} \E \left[\frac{||\bm{s}_{t+1} - \bm{x}||_2^2}{||\bm{x}||_2^2} \right].
\end{equation}
In the experiment, 
{ The pseudo inverse matrix is chosen as the matrix $W$ in OAMP to make the time complexity $O(N^2)$
in each iteration.
The divergence-free function of OAMP  in (\ref{oamp_s}) is based on the MMSE estimator (\ref{mmse_est}).}

From Fig. \ref{A_variance:1_M}, we can observe that TISTA provides the steepest  
NMSE curve among those algorithms in the first $12$ rounds. 
For example, OAMP and LISTA require $6$ and $10$ rounds, respectively, 
in order to achieve NMSE = $-30$ dB,  
whereas TISTA requires only $5$ rounds.
The NMSE curve of TISTA saturates at around $-42$ dB, at which 
TISTA and OAMP converge. This means that TISTA shows significantly faster convergence than AMP and LISTA in this setting.
TISTA also overwhelms OAMP in the NMSE performance.
TISTA has about $5.8$ dB and $4.0$ dB gains at $T=5$ and $7$ { compared with OAMP}, respectively.

In order to study the behavior of the learned trainable variables $\gamma_t$, we conducted the following experiments.
For a fixed sensing matrix $(\bm{A}_{i,j} \sim \mathcal{N}(0,1/M))$, 
we trained TISTA three times with distinct random number seeds.
The learned variables $\gamma_t$ (denoted by matrix 1--3) 
are shown in Fig. \ref{A_variance:1_M:g}.
The three sequences of learned parameters approximately coincide with each other.
Furthermore, the sequences have a zigzag shape, and the values of $\gamma_t$ lies in the range {from 1 to 10}.
{As for other trainable parameters, $\alpha^2$ is tuned to $3.68$-$3.71$ and $p$ is tuned to $0.08$-$0.09$.
Interestingly, the trained $\alpha$ becomes larger than the true value $1.0$ though $p$ does not change largely from the true value $0.1$.
Note that training these values improves the NMSE performance of TISTA, 
which suggests that the true values of parameters in the MMSE estimator are not always best for TISTA.}

\subsubsection{Large-scale problem}

\begin{figure}
        \begin{center}
          \includegraphics[width=0.95\linewidth]{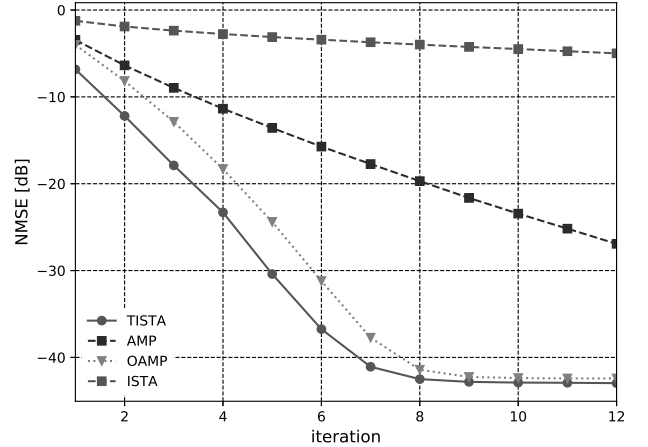}
          \caption{NMSE of TISTA and and other algorithms; {$N=5000, M = 2500$}, $p = 0.1, \bm{A}_{i,j} \sim \mathcal{N}(0, 1/M)$, SNR $= 40$ dB.}
          \label{A_variance:1_M:N5000}
        \end{center}
\end{figure}

As discussed in the previous section, the number of trainable variables 
of TISTA is considerably small. This feature enables us to handle large-scale problems.
Figure \ref{A_variance:1_M:N5000} shows the NMSEs for the cases of $(N, M) = (5000, 2500)$.
LISTA is omitted { from the comparison} because it is { computationally intractable to execute} in our environment.
We find that the NMSE performance of each algorithm are slightly better than that in the small system ($N=500$).
The gain of TISTA, however, is still large in this case.
In addition, TISTA saturates about $-43$ dB, which is $0.6$ dB lower than OAMP.
From these observations, we find that TISTA exhibits a good NMSE performance even in a large system.

\subsubsection{Running time}

\begin{figure}[bt]
        \begin{center}
          \includegraphics[width=0.95\linewidth]{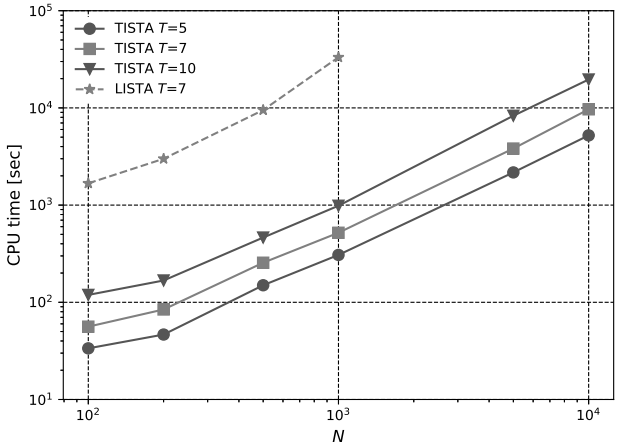}
          \caption{CPU time for learning and executing TISTA (solid lines) and LISTA (dashed line) as a function of $N$ with various $T$; $M/N=0.5$, SNR$=40$ dB.}
          \label{calc_time}
        \end{center}
\end{figure}

{ In order to demonstrate the scalability of TISTA explicitly, we show the CPU time 
required for training processes in Fig.~\ref{calc_time}.}
The CPU time is measured by a PC with Intel Xeon(R) CPU (3.6 GHz, 6 cores) and no GPUs.
It consists of the whole incremental training process up to $T$ layers and execution process of TISTA implemented by PyTorch 0.4.1.
In the experiment, we fix the rate $M/N$ to $0.5$ and SNR to 40 dB as the same setting with the previous experiments.
The results show that, in the case of $N=500$, TISTA is about 37 times faster than LISTA
 in addition to better NMSE performance as shown in Fig.~\ref{A_variance:1_M}.
We also find that TISTA has a notable scalability.
The CPU time of TISTA ($T=7$) for $N=10^4$ signals is nearly equal to that of LISTA ($T=7$) for $N=500$.
Simple linear regressions estimate that the CPU time roughly depends on $N^{1.2}$ and $T^{2.0}$.
These facts suggest that the small number of trainable parameters in TISTA enables its fast learning process for large problems.

\subsection{Gaussian sensing matrices with large variance}

\begin{figure}[bt]
        \begin{center}
          \includegraphics[width=0.95\linewidth]{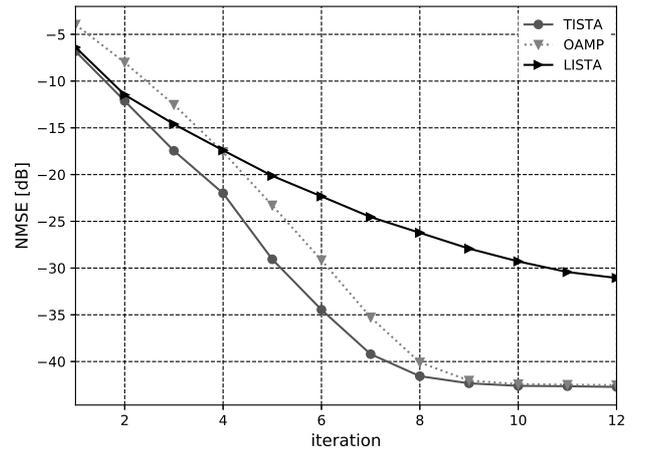}
          \caption{NMSE of {TISTA, OAMP, and LISTA};
          $\bm{A}_{i,j} \sim \mathcal{N}(0, 1)$, {$N = 500, M = 250$,}  SNR $= 40$ dB. 
          In this case, AMP cannot converge because the variance of the matrix components is too large.}
          \label{A_variance:1}
        \end{center}
\end{figure}

In the next experiment, we changed the variance of the sensing matrices to {a larger value}, 
i.e., each element in $\bm{A}$ follows $\mathcal{N}(0, 1)$ instead of $\mathcal{N}(0, 1/M)$.
{The trainable parameters of TISTA are $\{\gamma_t\}_{t=0}^{T-1}$, $\alpha$, and $p$.}
Figure \ref{A_variance:1} shows the NMSE curves of {TISTA, OAMP, and LISTA}.
Note that, under this condition,  
AMP does not perform well, i.e., AMP actually cannot converge at all, 
because the setting does not fit the required condition 
($\bm{A}_{i,j} \sim \mathcal{N}(0, 1/M)$) for achieving 
the guaranteed performance and the convergence of AMP.
As shown in Fig. \ref{A_variance:1},  
TISTA behaves soundly and shows
faster convergence than that of {OAMP and LISTA}.
This result suggests that TISTA 
is appreciably robust against the change of the variance.

\subsection{Binary matrix}

In this subsection, we will discuss the case in which 
the sensing matrices are binary, i.e.,  $\bm{A} \in \{\pm1\}^{M \times N}$.
Each entry of $\bm{A}$ is selected uniformly at random on $\{\pm1\}$.
This situation is closely related to multiuser detection in 
Coded Division Multiple Access (CDMA) \cite{reBP}.
\begin{figure}[bt]
        \begin{center}
          \includegraphics[width=0.95\linewidth]{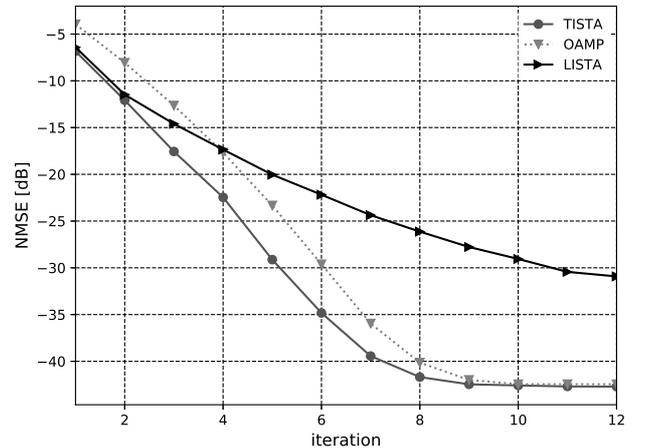}
          \caption{NMSE of {TISTA, OAMP, and LISTA}; $\bm{A}_{i,j}$ takes 
          a value in $\{\pm 1\}$ uniformly at random. {$N = 500, M = 250$,} SNR $= 40$ dB. AMP is not applicable in this case.}
          \label{A_binary}
        \end{center}
\end{figure}
Figure \ref{A_binary} shows the NMSE curves 
of {TISTA, OAMP, and LISTA} as a function of
iteration.
{As the previous subsections, TISTA trains $\{\gamma_t\}_{t=0}^{T-1}$, $\alpha$, and $p$.}
The NMSE curves of TISTA approximately coincide with those of the Gaussian sensing matrices. This result can be regarded as an 
evidence for the robustness of TISTA for non-Gaussian sensing matrices.

\subsection{Sensing matrices with a large condition number}

Regression problems regarding a matrix with a large condition number are difficult to solve in an accurate manner. The condition number $\kappa$ of a matrix is defined as the ratio of the
largest and smallest singular values, i.e.,
$\kappa = s_1/s_M$, where $s_1 \ge  s_2 \ge \cdots \ge s_M$ are the singular values of the matrix. 
In this subsection, we assess the performance of
TISTA for sensing matrices with a large condition number.
{In this subsection, the trainable parameters of TISTA are only $\{\gamma_t\}_{t=0}^{T-1}$
because it shows enough performance improvement.}

The setting for the experiments is as follows.
For a given condition number 
$\kappa$, we assume that the ratio $s_i / s_{i-1}$ is constant for each $i$
in order to fulfill $s_1/s_M = \kappa$ and $\tr{\bm{A} \bm{A}^T}=N$.
We first sample a matrix $\bm{G} \in \mathbb{R}^{M \times N}$, where each entry of $\bm{G}$ follows
an i.i.d. zero-mean Gaussian distribution with variance 1. 
The matrix $\bm{G}$ is then decomposed by singular value decomposition and we obtain    
$
\bm{G} = \bm{U} \bm{\Sigma} \bm{V}^T,
$
where $\bm{U} \in \mathbb{R}^{M \times M}$, $\bm{V} \in \mathbb{R}^{N \times N}$,
and $\bm{\Sigma} \in \mathbb{R}^{M \times N}$. 
From the set of singular values $s_1, \ldots, s_M$ satisfying 
the above conditions, $\bm{\Sigma}^*$ is defined by 
$
\bm{\Sigma}^* = (\bm{\Delta}\ \bm{O}), 
$
where the matrix $\bm{\Delta} =\mbox{diag}(s_1, \ldots, s_M)$, and $\bm{O}$ is the zero matrix.
A sensing matrix $\bm{A}$ with the condition number $\kappa$ is obtained by calculating 
$
	\bm{A} = \bm{U} \bm{\Sigma}^* \bm{V}^T.
$ 

\begin{figure}[bt]
        \begin{center}
          \includegraphics[width=0.95\linewidth]{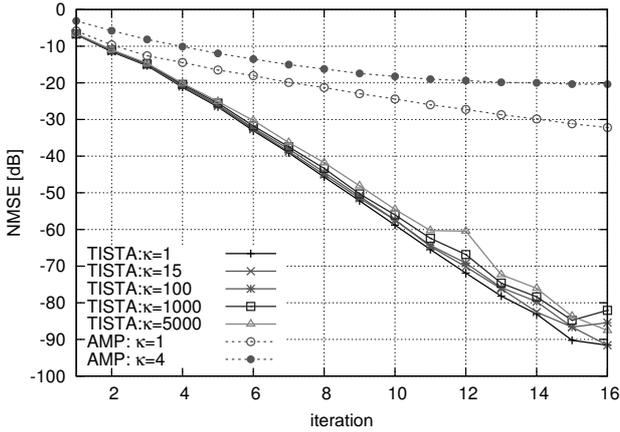}
          \caption{NMSE of TISTA and AMP; $\kappa$ represents the condition number. No observation noise ($\sigma^2 = 0$).}
          \label{A_cond:noiseless}
        \end{center}
\end{figure}

Figure \ref{A_cond:noiseless} shows the NMSE of TISTA and AMP without 
observation noise, i.e., $\sigma^2 = 0$. 
As shown in Fig. \ref{A_cond:noiseless}, there is almost no performance degradation in the NMSE even for a large condition number 
such as  $\kappa=5000$.
On the other hand, AMP converges up to $\kappa = 4$, but
the output diverges when $\kappa \ge 5$. 
These results indicate the robustness of TISTA with respect to sensing matrices with a large condition number in the noiseless case.

\begin{figure}[bt]
        \begin{center}
          \includegraphics[width=0.95\linewidth]{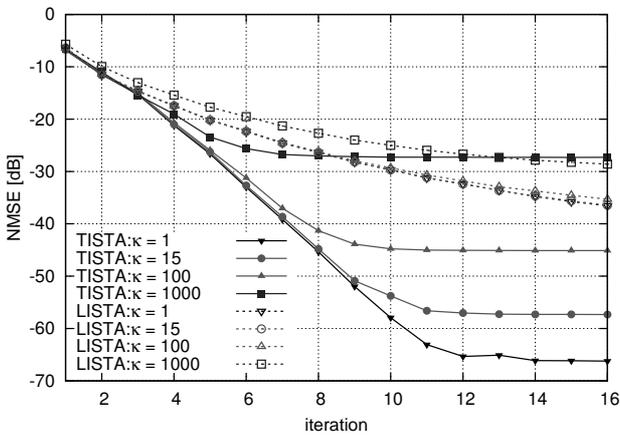}
           \caption{NMSE of TISTA and LISTA; $\kappa$ represents the condition number.
          SNR $= 60$ dB. }
          \label{A_condition}
        \end{center}
\end{figure}

Figure \ref{A_condition} shows the NMSE of TISTA and LISTA
when there are observation noises (SNR $= 60$ dB).
Compared with the NMSE curve of LISTA,  TISTA provides a much smaller 
NMSE in the cases of $\kappa = 1, 15, 100$.
However, in contrast to the noiseless case (Fig. \ref{A_cond:noiseless}), 
the NMSE performance of TISTA severely degrades as $\kappa$ increases.
This phenomenon can be considered as a consequence of the use of the pseudo inverse linear estimator $\bm{W}$, which tends to cause noise enhancement if the condition number is large.

\section{Hypothesis on zigzag shapes}

In the previous section, we observed that the trained values of $\{\gamma_t \}_{t=0}^{T-1}$ show
zigzag shapes that is not easy to interpret. The zigzag pattern yields the fast convergence 
property of TISTA and it should be a reasonable choice for accelerating its search processes.
 In this section, we try to provide a plausible hypothesis on the zigzag shapes.

We first consider a toy example for minimizing a quadratic function
$f(x_1, x_2) = x_1^2 + 10 x_2^2$ by using the gradient descent (GD) method.
The function is simple but the condition number regarding the problem is relatively large. 
This means that a naive GD method is not suitable for attaining fast convergence to the minimum point.
The main step of the GD method is the update of the search point as
\begin{equation}
\bm{s}_{t+1} = \bm{s}_{t} - \gamma \nabla f(\bm{s}_t)
\end{equation}
for $t = 1,2,\ldots, T$.
The parameter $\gamma$ is the step size parameter that significantly affects the behavior of the search process.
In this section, we assume that each element of the initial point $\bm{s}_1 = (s_{1,1}, s_{1,2})$ is chosen 
in the closed domain $[-10, 10]^2$ uniformly at random.

\begin{figure}[ht]
\begin{center}
\includegraphics[scale=0.8]{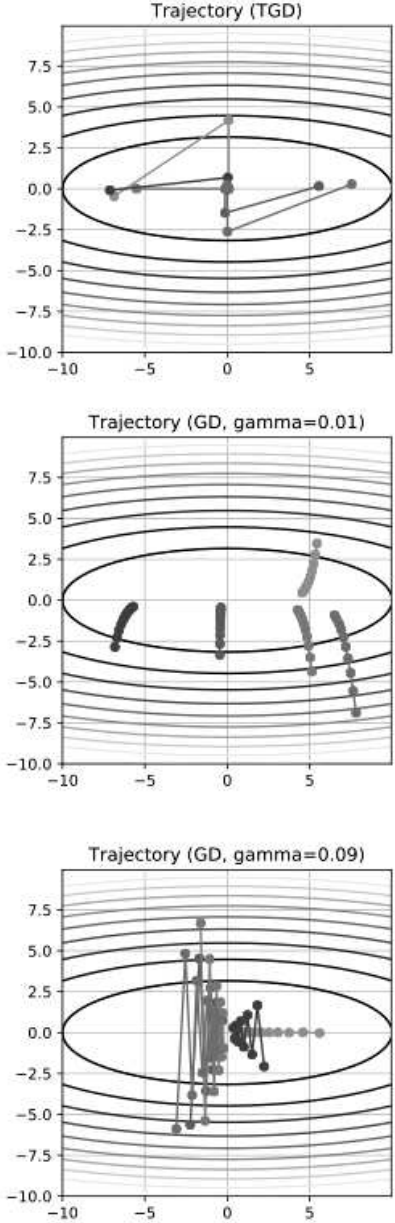}	
\caption{Trajectories of search points  (5 trials) in GD processes for $f(x_1, x_2) = x_1^2 + 10 x_2^2$: TGD (top), 
GD with $\gamma = 0.01$ (center),  GD
with $\gamma = 0.09$ (bottom). The optimal point is $(0, 0)$. The ovals are contour of the objective function.}
\label{trajectory}
\end{center}
\end{figure}

Figure \ref{trajectory} (center, bottom) shows typical minimization processes of the GD method.
A small step size (center) leads to considerably slow convergence but a large step size (bottom) 
induces oscillation behaviors 
that also slow down the convergence or lead to divergence. 

According to the idea of TISTA, i.e., embedding of trainable parameters, 
we can embed trainable parameters in the GD step as 
\begin{equation}
\bm{s}_{t+1} = \bm{s}_{t} - \gamma_t \nabla f(\bm{s}_t),
\end{equation}
where $\{\gamma_t\}_{t=1}^T$ is a set of trainable parameters.
The incremental training can be applied to train these parameters 
in order to accelerate the convergence. We call this method the trainable GD (TGD) hereafter.

\begin{figure}[htbp]
\begin{center}
\includegraphics[scale=0.6]{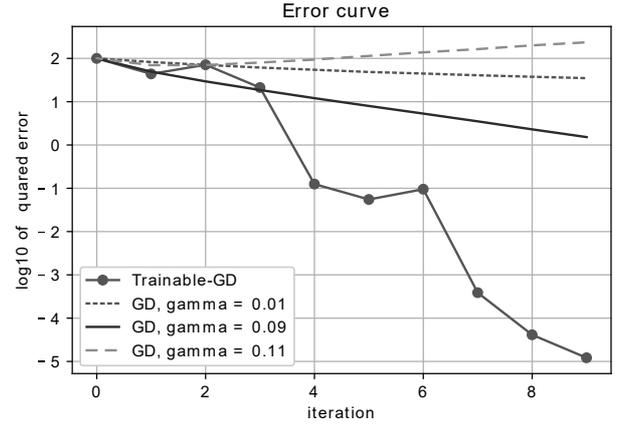}	
\caption{Averaged error curves of TGD and GD: The horizontal axis represents the number of iterations and 
the vertical axis represents the averaged error $\log_{10} ||\bm{s}_t - \bm{s}^*||_2^2$ where $\bm{s}_t$ is 
the search point after $t$ iterations, and $\bm{s}^*$ is the optimal solution. In the evaluation process, 
the outcomes of 10000 minimization trials with random starting points are averaged.}
\label{TGD_error}
\end{center}
\end{figure}

\begin{figure}[htbp]
\begin{center}
\includegraphics[scale=0.6]{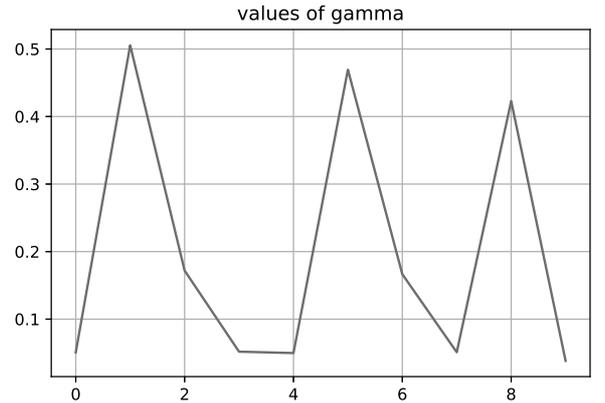}	
\caption{Trained values of $\gamma_i$: the details of the training is as follows. The incremental training with 
the mini-batch size 50 is used. In a generation, 500 mini-batches are processed. The optimizer is Adam with learning 
rate 0.001. }
\label{TGD_gamma}
\end{center}
\end{figure}

Figure \ref{TGD_error} shows the averaged error of TGD and GD as a function of the number of iterations.
TGD significantly  outperforms GD methods and provides much faster convergence. 
From the training process, TGD learns an appropriate strategy to yield fast convergence.
The trained values of $\{\gamma_t\}_{t=1}^T$ are plotted in Fig.~\ref{TGD_gamma}. 
We can observe a zigzag shape that represents the learned acceleration strategy for this problem. 
It is interesting to see that 
the behavior of the search point shown in Fig.~\ref{trajectory} (top)
is not similar to those of $\gamma = 0.01$ (center) nor $\gamma = 0.09$ (bottom).

Our hypothesis of the zigzag shapes is that a similar situation happens in signal recovery processes of TISTA as well.
The linear estimation step (\ref{lmse}) of TISTA  is closely related to the gradient descent step for the quadratic problem 
to minimize $||\bm{Ax} - \bm{y}||_2^2$, i.e., we have the exact gradient descent step by replacing $\bm{W}$ with $\bm{A}^T$.
If the quadratic problem is ill-conditioned or nearly ill-conditioned, the preferable strategy would be  
the {\em zigzag strategy} observed in Fig.~\ref{TGD_gamma} as well.
We still lack enough evidences to confirm the validity of the hypothesis and it should be confirmed in a future work.

\section{Sparse signal recovery for MNIST images}
In Sec. \ref{sec_eval}, we have seen results of the numerical experiments based on artificial sparse signals 
generated according to the i.i.d. Bernoulli-Gaussian prior model.
The feasibility of TISTA for sparse signals in the real world has not yet been clear
 because a real sparse signal may not follow the i.i.d. assumption.
In order to evaluate the performance of TISTA for non-i.i.d. signals,
we made experiments of sparse signal recovery based on the MNIST dataset.
The MNIST dataset is a dataset  including monochrome images of hand-written numerals and the corresponding labels.
Since most of pixels of an MNIST image is zero,  the MNIST dataset can be regarded as a dataset of sparse signals.
The goal of this section is to discuss the sparse signal recovery performance of TISTA for the MNIST dataset.

The details of the experiment is as follows.
An MNIST image  consists $28 \times 28 = 784$ pixels where
a pixel takes an integer value from $0$ to $255$.  We first normalize the pixel values to $[0, 1]$ 
and then rasterize the pixels as $784$-dimensional vectors.
In the following, we let $N = 784$ and $M = 392$. 
As a sensing matrix, we prepare a random matrix $\bm{A} \in \mathbb{R}^{M \times N}$
where each element in $A$ follows Gaussian distribution with zero mean and variance ${1}/{M}$.
We assume a noisy observation by the matrix $\bm{A}$ with the additive white Gaussian noise $\bm{w}$ with zero mean
and variance {$4 \times 10^{-4}$}, i.e., the received signal $\bm{y}$ is generated by $\bm{y}= \bm{A}\bm{x}+\bm{w}$.
As a sparse signal recovery algorithms, we compare TISTA with OAMP.
We choose the MMSE estimator (\ref{mmse_est}) for Bernoulli-Gaussian prior
as their MMSE functions because we assume that we have no knowledge on the prior PDF of the images.
We set the parameters of the prior to $\alpha=1$, $p=0.5$ for OAMP
while these parameters are trained from the dataset in TISTA.

The detail of the training processes is as follows.
In the training process of TISTA, as well as $\{\gamma_t\}_{t=0}^{T-1}$, the parameters $\alpha$ and $p$ are 
treated as trainable parameters. The size of mini-batch is set to $200$.
For a generation of incremental training, we used all the images in the MNIST training set (60000 images).
Adam optimizer with learning rate $0.005$ was used for training.

\begin{figure}
\begin{center}
\includegraphics[width=0.95\linewidth]{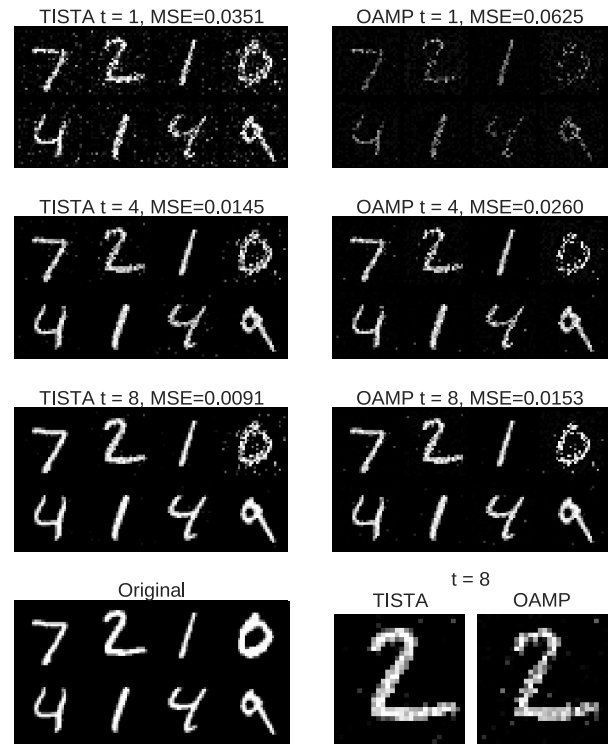}
\caption{Reconstructed images by TISTA (left column) and OAMP (right column). Parameters: $N = 784, M = 392$, 
$\bm{A}_{i,j} \sim \mathcal{N}(0, 1/M)$, {noise variance $4\times 10^{-4}$. The ``2'' images reconstructed by TISTA and OAMP with $t=8$
are shown in the right bottom for comparison.} }
\label{MNIST_comp}
\end{center}
\end{figure}

Figure \ref{MNIST_comp} shows the recovered images by TISTA (left column) and OAMP (right column)
with {$t = 1,4,8$} iterations. {These images are recovered from the same noisy observation of
the original image displayed on the left bottom.}
It can be observed that TISTA with $t = 8$ provides a reconstructed image considerably close to the 
original ({$\mathrm{MSE}=0.0091$}). The number ``0'' is not perfectly recovered because the original image is not so sparse and it 
affects the reconstruction quality.
The quality of the reconstructed images of TISTA evidently outperforms that of OAMP. For example,
{even with $t = 100$, the image reconstruction by OAMP ($\mathrm{MSE}=0.0148$) is worse than that by TISTA in terms of MSE.
In fact, we find that the reconstructed ``2'' by OAMP is not so crisp and clear compared with those of TISTA (right bottom of Fig. \ref{MNIST_comp}).
It implies that the training parameters $\alpha$ (trained value $1.59$) and $p$ (trained value $0.4$) positively affects the image reconstruction quality.
}

Moreover, comparing the images of $t = 1,4,8$, it can be confirmed that TISTA shows much faster convergence 
than OAMP. 
This tendency exactly coincides with the results reported in Section IV.

The result of this section strongly suggests that 
TISTA can be applied to sparse signal recovery problems based on the real data with non-i.i.d. sparse signals 
 if we have enough data to train the trainable parameters.

\section{Extensions}
In this section, we propose a few extensions of TISTA to treat a sensing matrix with nonzero-mean components or with a large condition number.
The numerical results show that the proposed extensions outperform the original TISTA in each situation without additional computational costs in the learning process.
{In this section, the trainable parameters of TISTA are only $\{\gamma_t\}_{t=0}^{T-1}$.}

 \subsection{Sensing matrices with nonzero-mean components}

In this subsection, we propose an extension of TISTA for a sensing matrix with nonzero-mean components.
It is known that, e.g., generalized AMP~\cite{GAMP} (GAMP), which is constructed for 
zero-mean Gaussian random matrices, fails to converge to a fixed point
when a sensing matrix consists of nonzero-mean components~\cite{IC1}.
To overcome this difficulty, Vila {et al.} proposed a variant of GAMP with damping of messages and mean removal from a sensing matrix and signals~\cite{IC2}.
 Following these advances in AMP, we apply a mean removal technique 
 to TISTA to improve its performance for large nonzero-mean sensing matrices.

Let us consider {\em TISTA-MR}, TISTA with the mean removal technique.
We assume that the sensing matrix $\bm{A}$ is generated according to 
the Gaussian distribution $\mathcal{N}(\mu_{\bm{A}},\sigma^2)$ with a nonzero mean $\mu_{\bm{A}}$.
In fact, without any modifications, TISTA shows poor performance as $\mu_{\bm{A}}$ increases.
The simplest extension involves the use of a modified sensing matrix $\bm{A}'=(\bm{A}'_{i,j})$, where 
$\bm{A}'_{i,j}= \bm{A}_{i,j} -\mu_{\bm{A}}$ instead of an original sensing matrix $\bm{A}=(\bm{A}_{i,j})$.
The modified recursion formula of TISTA is then written as follows:
\begin{eqnarray}
\bm{u}_t &=& \bm{y}-\bm{A}'\bm{s}_t \label{mean},\\
\bm{r}_t &=& \bm{s}_t + \gamma_t \bm{W}' \left(\bm{u}_t - \frac{1}{M}\bm{1}_{M}^{T} \bm{u}_t \bm{1}_{M}\right) \label{lmse_m}\\
\bm{s}_{t+1} &=& \mmse {\bm{r}_t; \tau_t^2} \label{mmse_m}\\
v_t^2 &=& \max \left\{ \frac{||\bm{u}_t - \frac{1}{M}\bm{1}_{M}^{T} \bm{u}_t \bm{1}_{M}||_2^2 - M \sigma^2}{\tr{\bm{A}'^T \bm{A}'}}, \epsilon \right\}\label{v_t_m}\\
\tau_t^2 &=& \frac{v_t^2}{N} (N + (\gamma^2_t-2\gamma^2_t) M )   \nonumber\\
&+& \frac{\gamma_t^2\sigma^2}{N} \tr{\bm{W}' \bm{W}'^T}, \label{tau_m}
\end{eqnarray}
where $\bm{1}_{M}= (1,1,\dots,1)^{T}$ is an $M$-dimensional vector, the elements of which are 1s, and matrix $\bm{W}'$ is the pseudo inverse matrix of $\bm{A}'$.
In the formula, $\bm{r}_t$ is calculated via $\bm{u}_t - M^{-1}\bm{1}_{M}^{T} \bm{u}_t \bm{1}_{M}$ 
to remove the mean of $\bm{u}_t$.
These modifications enable the performance of TISTA-MR to be improved because it attempts to recover a sparse signal with a modified sensing matrix,
 the components of which have sufficiently small means.
Note that further performance improvement may be achieved when we use a modified sensing matrix for which the means of rows and columns are expected to be zero, as in~\cite{IC2}.

\begin{figure}[bt]
        \begin{center}
          \includegraphics[width=0.95\linewidth]{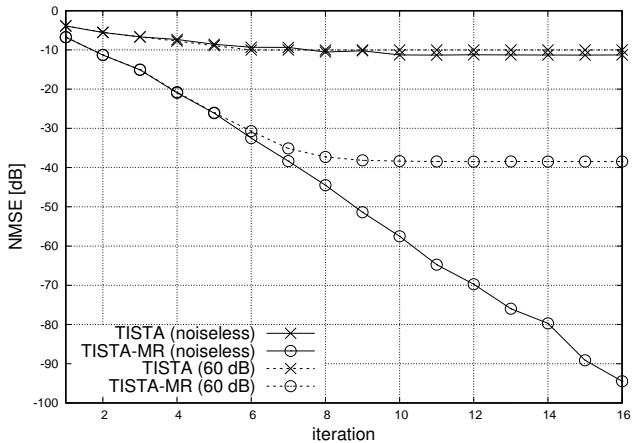}
          \caption{NMSE of the original TISTA (cross marks) and TISTA-MR (circles) with mean removal; $A_{i,j} \sim \mathcal{N}(1, 1/M)$, $N=500$, $M=250$. No observation noise ($\sigma^2=0$)
          and SNR = $60$ dB cases.}
          \label{A_mean:1}
        \end{center}
\end{figure}

Figure~\ref{A_mean:1} shows the NMSE of the original TISTA and {TISTA-MR for noiseless case} in the case of noiseless observation and SNR = $60$ dB.
Each element of a sensing matrix $\bm{A}$ is generated from $\mathcal{N}(1,1/M)$, where the original AMP has difficulty in convergence.
{TISTA-MR} outperforms the original TISTA for which the NMSE saturates around $-10$ dB in both cases.
In the case of SNR = $60$ dB, {TISTA-MR} scores $-38$ dB in the NMSE  with about $28$ dB gain against TISTA when $T=10$.
These numerical results indicate that {TISTA-MR} based on mean removal gives drastically improved signal recovery performance without increasing the time complexity.

 \subsection{Sensing matrices with a large condition number}

As discussed in the previous section, TISTA exhibits a non-negligible performance degradation
(except for the noiseless case) when the condition number of the sensing matrix is large.
In this subsection, we present a method for improving the sparse recovery performance of TISTA in such a case by using an LMMSE matrix as a linear estimator.
A naive approach to suppress the noise enhancement in linear estimation 
is to use the LMMSE matrix
\begin{equation} \label{MMSE_W}
\bm{W}_t = v_t^2 \bm{A}^T(v_t^2 \bm{A} \bm{A}^T + \sigma^2 \bm{I})^{-1}
\end{equation}
as a linear estimator in TISTA recursions. 
Note that the error variance $v_t^2$ is calculated in a recursive calculation process of TISTA.
Ma and Ping \cite{OAMP} took this approach in their OAMP experiments. 
A drawback of this approach is that it is necessary to calculate an $M \times M$ matrix inversion in (\ref{MMSE_W}) for each iteration, which requires $O(M^3)$ time for an iteration.
In order to avoid the matrix inversion for each iteration,
we use a simple ad-hoc solution, and define the matrix $\bm{W}$ as 
\begin{equation}\label{W_improve}
\bm{W} = \bm{A}^T(\bm{A} \bm{A}^T + \beta \bm{I})^{-1},
\end{equation}
where $\beta$ is a real constant.
We call TISTA with~(\ref{W_improve}) {\em TISTA-LMMSE}.
This is the only difference from the original TISTA using the pseudo 
inverse matrix of $\bm{A}$ as $\bm{W}$. The term $\beta \bm{I}$
can decrease the condition number of $\bm{W}$ and 
prevents noise enhancement.
{ Matrix inversion is necessary only once at the beginning of a recovery process. }
Thus, the required time complexity of TISTA-LMMSE is the same as that of the original TISTA.
The parameter $\beta$ is determined to minimize the value of the NMSE after training.

\begin{figure}
        \begin{center}
          \includegraphics[width=0.95\linewidth]{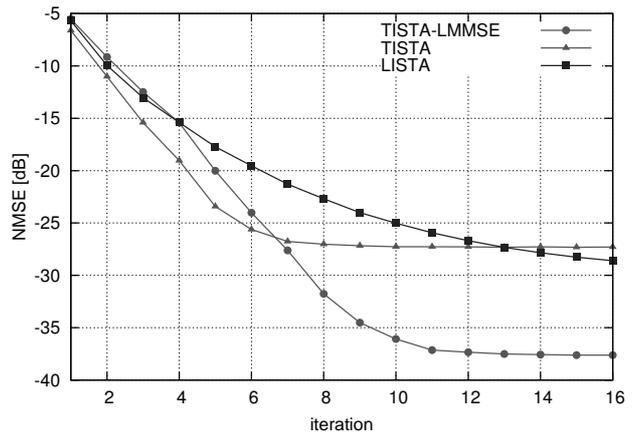}
          \caption{NMSE of LISTA, the original TISTA, and TISTA-LMMSE with (\ref{W_improve}) 
          ($\beta = 5.0 \times 10^{-4}$); 
          condition number $\kappa=1000$,  SNR $= 60$ dB.}
          \label{A_condition:improve}
        \end{center}
\end{figure}

Figure \ref{A_condition:improve} shows the NMSE curves for the case of $\kappa = 1000$, which includes the NMSE curve 
of TISTA-LMMSE with (\ref{W_improve}).
In TISTA-LMMSE, we used the parameter $\beta=5.0 \times 10^{-4}$. From Fig. \ref{A_condition:improve}, we can confirm that 
TISTA-LMMSE exhibits much better NMSE performance as compared with the original TISTA using the pseudo inverse matrix in the linear estimator.
This example shows that this simple ad-hoc approach is fairly effective without additional cost.

\section{Conclusion}

The crucial feature of TISTA is that it includes adjustable variables 
which can be tuned by standard deep learning techniques. 
The number of trainable variables of TISTA is equal to the number of iterative rounds and is 
much smaller than those of the known learnable sparse signal recovery algorithms \cite{LISTA, LAMP, Borgerding}.
This feature leads to the highly stable and fast training processes of TISTA.
Computer experiments indicate that TISTA is applicable to various classes of sensing matrices 
such as Gaussian matrices, binary matrices, and matrices with large condition numbers.
Furthermore, numerical results demonstrate that TISTA shows significantly faster convergence 
than AMP or LISTA in many cases and remarkably large gains compared to OAMP. 
{ The experimental results on the MNIST image set imply
that TISTA is also applicable for non-i.i.d. sparse signals in the real world. }
{In summary, TISTA achieves remarkable performance improvement for artificial data and promising flexibility to real data
with fast learning process, high stability, and high scalability using a quite simple architecture.}

{For a future plan, by} replacing the MMSE shrinkage, we can expect that TISTA is also applicable to non-sparse signal recovery problems 
such as detection of BPSK signals in overloaded MIMO systems \cite{IW-SOAV2}.
{ Another possibility is to replace the MMSE shrinkage function 
with a small neural network that can learn an appropriate shrinkage function matched to the prior of the sparse signals.
This change could significantly broaden the target of TISTA.}

 \section*{Acknowledgement}
The authors would like to thank the anonymous reviewers of ICC2018 
{ and IEEE Transactions on Signal Processing} for their constructive comments.
The present study was supported by JSPS Grant-in-Aid for Scientific Research (B) Grant Number 16H02878 (TW)
and Grant-in-Aid for Young Scientists (Start-up) Grant Number 17H06758 (ST).
The last author is grateful to Dr. Keigo Takeuchi {for the inspiring seminar at Nagoya Institute of Technology}.


\begin{thebibliography}{99}
\bibitem{CS1} D. L. Donoho, 
``Compressed sensing,'' 
IEEE Trans. Inf. Theory, vol. 52, no. 4, pp. 1289-1306, Apr. 2006.

\bibitem{CS2} E. J. Candes and T. Tao,
 ``Near-optimal signal recovery from random projections: Universal encoding strategies?'' 
IEEE Trans. Inf. Theory, vol. 52, no. 12, pp. 5406-5425, Dec. 2006.

\bibitem{Alg_survey} Z. Zhang, Y. Xu, J. Yang, X. Li, and D. Zhang, 
``A survey of sparse representation: Algorithms and applications,''
 IEEE Access, vol. 3, pp. 490-530, May. 2015.

\bibitem{LASSO} R. Tibshirani,
``Regression shrinkage and selection via the lasso,''
J. Royal Stat. Society, Series B, vol. 58, pp. 267-288, 1996.

\bibitem{MP} G. Davis, S. Mallat, and M. Avellaneda, ``Adaptive greedy approximations,'' 
Constructive Approximation, vol. 13, no. 1, pp. 57-98, Mar. 1997.

\bibitem{LARS} B. Efron, T. Hastie, I. Johnstone, and R. Tibshirani, 
``Least angle regression,''
Ann. Stat., vol. 32, no. 2, pp. 407-499, Apr. 2004.

\bibitem{CDA} T. T. Wu and K. Lange,
 ``Coordinate descent algorithms for lasso penalized regression,''
Ann. Appl. Stat., vol. 2, no. 1, pp. 224-244, 2008.

\bibitem{ISTA} A. Chambolle, R. A. DeVore, N. Lee, and B. J. Lucier,
``Nonlinear wavelet image processing: Variational problems, compression, and noise removal through wavelet shrinkage,''
IEEE Trans. Image Process., vol. 7, no. 3, pp. 319-335, Mar, 1998.

\bibitem{ISTA2} I. Daubechies, M. Defrise, and C. De Mol,
``An iterative thresholding algorithm for linear inverse problems with a sparsity constraint,''
Comm. Pure and Appl. Math., vol. 57, no. 11, pp. 1413-1457, Aug. 2004.

\bibitem{Prox} N. Parikh and S. Boyd, 
``Proximal algorithms,''
Foundations and Trends in Optimization, vol. 1, no. 3, pp. 123-231, 2014.


\bibitem{reBP} Y. Kabashima, 
``A CDMA multiuser detection algorithm on the basis of belief propagation,'' 
J. Phys. A: Math. Gen., vol. 36 pp. 11111-11121, Oct. 2003.

\bibitem{AMP} D. L. Donoho, A. Maleki, and A. Montanari,
``Message-passing algorithms for compressed sensing,''
Proceedings of the National Academy of Sciences, vol. 106, no. 45, pp. 18914-18919, Nov. 2009.

\bibitem{SE1} D. L. Donoho, A. Maleki, and A. Montanari,
``Message passing algorithms for compressed sensing: I. Motivation and construction,''
\textit{IEEE Information Theory Workshop 2010}, pp. 1-5, Jan. 2010.

\bibitem{SE2} M. Bayati and A. Montanari, 
``The dynamics of message passing on dense graphs, with applications to compressed sensing,"
IEEE Trans. Inf. Theory, vol. 57, no. 2, pp. 764-785, Jan. 2011.

\bibitem{IC1} F. Caltagirone, L. Zdeborova, and F. Krzakala, 
``On convergence of approximate message passing,''
\textit{2014 IEEE Int. Symp. Inf. Theory}, Jun. 2014, pp. 1812-1816.

\bibitem{IC2} J. Vila, P. Schniter, S. Rangan, F. Krzakala, and L. Zdeborova, 
``Adaptive damping and mean removal for the generalized approximate message passing algorithm,''
 \textit{2015 IEEE International Conference on Acoustics, Speech and Signal Processing}, Apr. 2015, pp. 2021-2025.

\bibitem{OAMP} J. Ma and L. Ping,
``Orthogonal AMP,''
IEEE Access, vol. 5, pp. 2020-2033, Jan. 2017.

\bibitem{VAMP} S. Rangan, P. Schniter, and A. K. Fletcher, 
``Vector approximate message passing,''
 \textit{2017 IEEE Int. Symp. Inf. Theory}, Jun. 2017, pp. 1588-1592.

\bibitem{Takeuchi} K. Takeuchi, 
``Rigorous dynamics of expectation-propagation-based signal recovery from unitarily invariant measurements,''
 \textit{2017 IEEE Int. Symp. Inf. Theory}, Jun. 2017, pp. 501-505.


\bibitem{DNN1} K. Fukushima, 
``Neocognitron: A self-organizing neural network model for a mechanism of pattern recognition unaffected by shift in position,''
Bio. Cybern., vol. 36, no. 4, pp. 193-202, 1980.

\bibitem{DNN2} M. Riesenhuber and T. Poggio, 
``Hierarchical models of object recognition in cortex,'' 
Nature Neuroscience, vol. 2, no. 11, pp. 1019-1025, Nov. 1999.

\bibitem{Image1} G. E. Hinton, R. R. Salakhutdinov,
``Reducing the dimensionality of data with neural networks,''
Science, vol. 313, no. 5786, pp. 504-507, Jun. 2006.

\bibitem{Image2} A. Krizhevsky, I. Sutskever, G. E. Hinton, 
``Imagenet classification with deep convolutional neural networks.''
 \textit{Advances in Neural Inf. Process. Sys. 2012}, pp. 1097-1105, Dec. 2012.

\bibitem{Speech1} G. Hinton et al.,
``Deep neural networks for acoustic modeling in speech recognition: The shared views of four research groups,''
 IEEE Signal Processing Magazine, vol. 29, no. 6, pp. 82-97, Nov. 2012.

\bibitem{Speech2} G. E. Dahl, D. Yu, L. Deng and A. Acero,
"Context-dependent pre-trained deep neural networks for large-vocabulary speech recognition," 
 IEEE Trans. Audio, Speech, Lang. Process., vol. 20, no. 1, pp. 30-42, Jan. 2012.

\bibitem{Robo1} R. Hadsell, A. Erkan, P. Sermanet, M. Scoffier, U. Muller and Y. LeCun, 
``Deep belief net learning in a long-range vision system for autonomous off-road driving,''
 \textit{2008 IEEE/RSJ International Conference on Intelligent Robots and Systems}, Sep. 2008, pp. 628-633.

\bibitem{Com1} B. Aazhang, B. P. Paris and G. C. Orsak, 
``Neural networks for multiuser detection in code-division multiple-access communications,'' 
 IEEE Trans. Comm., vol. 40, no. 7, pp. 1212-1222, Jul. 1992.

\bibitem{Com2} E. Nachmani, Y. Be\'ery and D. Burshtein, 
``Learning to decode linear codes using deep learning,''
\textit{2016 54th Annual Allerton Conf. Comm., Control, and Computing}, 2016, pp. 341-346.

\bibitem{Com3} T. O'Shea and J. Hoydis, 
``An introduction to deep learning for the physical layer,''
IEEE Trans. Cog. Comm. Net., vol. 3, no. 4, pp. 563-575, Dec. 2017.

\bibitem{SGD} Y. A. LeCun, L. Bottou, G. B. Orr, and K. R. M\"uller,
``Efficient backprop,'' 
in \textit{Neural networks: Tricks of the trade},
G. B. Orr and K. R. M\"uller, Eds. Springer-Verlag, London, UK, 1998, pp. 9-50. 

\bibitem{BackProp} D. E. Rumelhart, G. E. Hinton, and R. J. Williams, 
``Learning representations by back-propagating errors,''
Nature, vol. 323, no. 6088, pp. 533-536, Oct. 1986.


\bibitem{LISTA} K. Gregor, and Y. LeCun,
``Learning fast approximations of sparse coding,''
 \textit{Proc. 27th Int. Conf. Machine Learning}, pp. 399-406, 2010.


\bibitem{LAMP}
M. Borgerding and P. Schniter, 
``Onsager-corrected deep learning for sparse linear inverse problems,''
\textit{2016 IEEE Global Conf. Signal and Inf. Process. (GlobalSIP)}, Washington, DC, Dec. 2016, pp. 227-231.


\bibitem{Borgerding}
M. Borgerding, P. Schniter, and S. Rangan,
``AMP-inspired deep networks for sparse linear inverse problems, ''
 IEEE Trans, Sig. Process. vol 65, no. 16, pp. 4293-4308 Aug. 2017.


\bibitem{Zhang}
J. Zhang and B. Ghanem, 
``ISTA-Net: Iterative shrinkage-thresholding algorithm 
inspired deep network for image compressive sensing,''
arXiv:1706.07929v1, 2017.


\bibitem{FISTA} A. Beck, and M. Teboulle,
``A fast iterative shrinkage-thresholding algorithm for linear inverse problems,''
SIAM J Imaging Sciences, vol. 2, no. 1, pp. 183-202, 2009.

\bibitem{TwIST} J. M. Bioucas-Dias and M. A. T. Figueiredo,
``A new TwIST: Two-step iterative shrinkage/thresholding algorithms for image restoration,''
 IEEE Trans. Image Process., vol. 16, no. 12, pp. 2992-3004, Dec. 2007.


\bibitem{EGAMP} J. Vila and P. Schniter,
``Expectation-maximization gaussian-mixture approximate message passing,''
IEEE Trans. Signal Process., vol. 61, no. 19, pp. 4658-4672, Oct. 2013.
%



\bibitem{Adam} D. P. Kingma and  J. L. Ba,
``Adam: A method for stochastic optimization,''
arXiv:1412.6980, 2014.

\bibitem{alpha} A. Montanari,
``Graphical models concepts in compressed sensing,''
in \textit{Compressed sensing: Theory and applications}, Cambridge University Press, Cambridge,
 pp. 394-438, 2012.

\bibitem{tens}
``TensorFlow: Large-scale machine learning on heterogeneous systems,''
\url{http://tensorflow.org/}
2015. Software available from tensorflow.org.

\bibitem{Pytorch}
A. Paszke et al., ``Automatic differentiation in PyTorch,'' \textit{31st Conf. Neural Inf. Process. Syst.}, pp. 1--4, 2017.
Software available from pytorch.org.

\bibitem{git}
\url{https://github.com/mborgerding/onsager_deep_learning/blob/master/README.md}

\bibitem{MMSE} P. Schniter, L. C. Potter and J. Ziniel,
``Fast bayesian matching pursuit,''
\textit{2008 Information Theory and Applications Workshop}, Jan. 2008, pp. 326-333.

\bibitem{Gribonval}
R. Gribonval
``Should penalized least squares regression be interpreted as maximum a posteriori estimation?, ''
IEEE Trans. Sig. Process., vol.59, no.5, pp. 2405-2410,  May. 2011.

\bibitem{Kazerouni}
A. Kazerouni, U. S. Kamilov, E. Bostan, and M. Unser, 
``Bayesian denoising: From MAP to MMSE using consistent cycle spinning, ''
IEEE Signal Process. Lett., vol. 20, no. 3, pp. 249-252, Mar. 2013.

\bibitem{GAMP} S. Rangan, 
``Generalized  approximate  message  passing  for estimation  with  random  linear  mixing,''
\textit{IEEE  Int. Symp. Inf. Theory}, Aug. 2011, pp. 2168-2172.



\bibitem{IW-SOAV2} R. Hayakawa and K. Hayashi, ``Convex optimization-based signal detection for massive overloaded MIMO systems,'' in IEEE Trans. Wireless Comm., vol. 16, no. 11, pp. 7080-7091, Nov. 2017.


\end{thebibliography}
\end{document}